\documentclass[12pt]{article}

\usepackage{amsfonts}
\usepackage{bbm}
\usepackage{bm}
\usepackage{mathrsfs}
\usepackage{slashed}
\usepackage{color}
\usepackage{geometry}             % See geometry.pdf to learn the layout options. There are lots.
\usepackage{graphicx}
\usepackage{amssymb}
\usepackage{amsmath}
\usepackage{epstopdf}
\usepackage{comment}

\usepackage[hyperindex=true,
          pdfstartview=FitH,
          bookmarksnumbered=true,
          bookmarksopen=true,
          citecolor=blue,
          linkcolor=blue,
          colorlinks=true,
          pdfborder=001,
          unicode]{hyperref}

\allowdisplaybreaks

\newcommand{\rd}[1]{\mathrm{d}#1}

\begin{document}

\title{Work and work-energy theorem in curved spacetime}

\author{Shaofan Liu and Liu Zhao\thanks{Correspondence author.}\\
School of Physics, Nankai University, Tianjin 300071, China
\\
{\em email}: \href{mailto:2120190132@mail.nankai.edu.cn}
{2120190132@mail.nankai.edu.cn}
and
\href{mailto:lzhao@nankai.edu.cn}{lzhao@nankai.edu.cn}
}
\date{}
\maketitle

\begin{abstract}
The definitions of gravitational work as well as work done 
by the total external force on a massive
probe particle moving in generic spacetime backgrounds are proposed. 
These definitions are given in the form of scalar integrals and thus, are 
independent of coordinate choices. However, the dependence on the choice 
of observer field is essential and inevitable. The definitions are
checked in the case of Minkowski, Schwarzschild, Reisner-Nordstr\"om and Kerr-Newman
spacetimes and agreements with Newtonian mechanical definitions are verified 
in the slow motion or the far field limit.
\end{abstract}

\section{Introduction}

In an attempt to extend classical thermodynamics and 
statistical physics into curved spacetime, we encounter the 
following problem: what is the work and heat in curved spacetime? 
Thinking a little bit further, we found that even on the microscopic
level (i.e. on the level of mechanics rather than statistical physics), 
the concept of work done by a force is still not clearly defined in the 
context of relativity. Discussions about the motion of a test particle in 
curved spacetimes can be easily found in most standard textbooks about general 
relativity, e.g. Section 4.3 in \cite{Wald}, however no clear and systematic definition 
of work done on a particle has been mentioned in any textbook on general relativity.

In Newtonian mechanics, the work done by a force exerted on a particle is one of the essential 
concepts defined as
\begin{align}
W=\int_C \bm{F}\cdot\rd{\bm{x}},
\end{align}
where $\bm{x}$ denotes the displacement along the path $C$ and the force 
$\bm{F}$ is in general dependent on $\bm{x}$: $\bm{F}=\bm{F}(\bm{x})$. 
This definition can also be recast in a number of different forms, e.g.
\begin{align}
W=\int_{t_i}^{t_f} \bm{F}\cdot \bm{v}\rd{t}
= \int_{t_i}^{t_f} m \bm{a}\cdot \bm{v}\rd{t},
\label{2}
\end{align}
due to the second law of Newtonian mechanics, $\bm{F}=m\bm{a}$, with $m$ the mass of the particle. 
For conservative forces, the work is path independent, because
\begin{align}
W=\int_C \bm{F}\cdot\rd{\bm{x}}
= -\int_{\bm{x}(t_i)}^{\bm{x}(t_f)} \nabla U(\bm{x})\cdot\rd{\bm{x}}
=U[\bm{x}(t_i)]-U[\bm{x}(t_f)].
\label{3}
\end{align}
In particular, the work of gravity exerted by a mass $M$ on another mass 
$m$ is 
\begin{align}
W=-\int_{\bm{r}(t_i)}^{\bm{r}(t_f)}\frac{GMm}{r^3}\bm{r}\cdot \rd{\bm{r}}
=GMm\left(\frac{1}{r(t_f)}-\frac{1}{r(t_i)}\right).
\label{Ng}
\end{align}
  %When the frame is noninertial, we can exert an "inertial force" so that the descriptions of work still hold, espetially for observers comoving with the mass $m$, the displacement $\rd{\bm{r}}$ is always zero and hence the work vanishes.
The above results only hold in inertial frames. For noninertial frames, however, 
we have to include a fictitious force $\boldsymbol{F}_I$ in order that Newton's 
second law still holds:
\begin{equation*}
\boldsymbol{F} + \boldsymbol{F}_I = m \boldsymbol{a}, \qquad \boldsymbol{F}_I \equiv - m \boldsymbol{a}_I.
\end{equation*}
where $\boldsymbol{a}_I$ is acceleration of the noninertial frame. The 
force $\boldsymbol{F}_I$ is generally called a pseudo force, or inertial force 
(see e.g. \cite{Berkly}). 
As the particle moves, nonzero fictitious work can be done on the particle by 
the inertial force. In the case of a uniform gravitational field with 
gravitational acceleration $\boldsymbol{g}$, the inertial force in the frame 
of a freely falling elevator is $-m\boldsymbol{g}$, so that the total
force exerted on a freely falling particle in this frame is zero.

In the context of general relativity, the concept of work becomes quite 
obscure. This obscurity comes into reality for a number of reasons. 
First of all, general relativity requires a covariant formalism.  
One can introduce a general covariant analogue of the second law of 
Newtonian mechanics, i.e.
\[
F^\mu = m a^\mu, \quad a^\mu\equiv u^\nu\nabla_\nu u^\mu.
\]
However, since the proper acceleration $a^\mu$ is always spacelike 
and is normal to the particle's proper velocity $u^\mu$,
\[
a^\mu u_\mu =0,
\]
the naive covariantization 
\begin{align}
W=\int_{t_i}^{t_f} \bm{F}\cdot \bm{v}\rd{t}\quad\longrightarrow \quad
W=\int_{\tau_i}^{\tau_f} F^\mu u_\mu \rd{\tau} 
\label{cov}
\end{align}
of eq.\eqref{2} does not work, because
\[
\int_{\tau_i}^{\tau_f} F^\mu u_\mu \rd{\tau}
= \int_{\tau_i}^{\tau_f} m a^\mu u_\mu\rd{\tau}=0.
\]
where $\tau$ is the particle's proper time, which is a natural 
parameter for the worldline of the particle. Secondly, as a physically 
measurable quantity, the definition of work 
must involve observer dependence, but the naive correspondence \eqref{cov}
clearly does not accommodate this information. 
The dependence of Newton's laws on inertial frames apparently violates the principles 
of relativity as no priority should be given to the inertial observers in a relativistic 
theory, and the measurement of work for all observers needs to be treated on equal 
footing. Lastly, when considering 
the work exerted by gravity, we encounter the most severe problem 
because gravity in general relativity is not viewed as a force
but rather is encoded in the spacetime geometry.

Discussions about observers in curved spacetimes have a long history. For examples, 
in 1909 Erenfest considered the Lorentz contraction of a rigid body observed by 
an observer at rest \cite{Erenf}, Doughty analyzed the proper acceleration of a 
static observer near the horizon of a black hole \cite{Doughty}, Crawford and his 
coworkers discussed generalized observer sets and measurements of velocity \cite{Crawford}. 
There are also discussions about dependence of quasi-local energy on 
observers \cite{Booth,Yu}.  Dahia and da Silva considered the relation between 
observers in curved spacetimes and those in Minkowski spacetime \cite{Dahia}. 
Berezin and Victor discussed the possibility to build up a set of static observers 
outside a Schwarzschild black hole \cite{Berezin}. Refs.\cite{Booth,Yu} also studied 
the work and binding energy associated to a massive shell \cite{Booth,Yu}. 
However no previous literature has provided a clear definition of work done on a 
moving relativistic particle.

In this paper, we propose a relativistic invariant definition of gravitational 
work as well as work done by the total external forces exerted on a probe particle 
in a generic curved spacetime. The crucial idea comes from the relativistic 
work-energy theorem, which asserts that the change in the energy of the particle 
is the sum of works done by the total external force and by gravity. 
In differential form, this means that the proper time rate of the energy
of the particle equals the sum of the local power of the total external force 
and that of gravity. It is crucial to notice that, in order to define works 
along a segment of the particle's worldline, a single observer is insufficient. 
Rather, a smooth observer field is a necessary input. 
Our definitions are checked in the case of several well-known spacetimes and the 
correctness in the non-relativistic limit is verified satisfactorily.

\section{The relativistic definition of work}

In this section we analyze gravitational work as well as work done by the total 
external force separately, and then add them up to get the total work. In order 
to define work in a curved spacetime, we need to analyze the local power measured 
by an on-the-spot observer as a preparation, and then integrate the power along 
a segment of the particle's worldline. This implies the inclusion of 
an observer field in the definition of work. We shall see that 
the expression of the local power of gravity contains the directional derivative 
of the proper velocity of the observer field along the tangent direction 
of the particle's worldline, thus the observer field needs to 
be smoothly distributed at least along the segment of the particle's worldline 
of our interest.

\subsection{Decomposition of the particle's proper velocity}

Consider a probe particle of mass $m$ moving in a generic spacetime $\mathcal{M}$ 
endowed with a metric $g_{\mu\nu}$ with mostly positive signature. A crucial 
quantity to be used throughout this paper is the energy $E(\mathcal{O})$ of the 
particle measured by a generic observer $\mathcal{O}$. 
The observer $\mathcal{O}$ is characterized by a future-directed 
timelike curve with tangent vector $Z^\mu$ normalized as $Z^\mu Z_\mu=-c^2$. 
Since parallel transport is curve dependent in curved spacetimes, there is 
generally no way for an observer to define the energy of a distant 
particle \cite{Wald}. Suppose that the worldline of the observer 
intersects with the worldline of the particle at the event $P$. Then the the energy of 
the particle at $P$ is defined as \cite{Wald}
\begin{equation}\label{engy}
E(\mathcal{O})  \equiv -Z^\mu p_\mu    = -m Z^\mu u_\mu .
\end{equation}
Introducing a scalar object
\begin{equation}\label{gamma}
\gamma   \equiv -\frac{1}{c^2} Z^\mu u_\mu  ,
\end{equation}
the energy can be rewritten as
\begin{equation}\label{E=gam}
E(\mathcal{O})   = \gamma   m c^2.
\end{equation}
It is always possible to make an orthogonal decomposition of $u^\mu$ with respect to 
$Z^\mu$, i.e.
\begin{equation}\label{dec u}
u^\mu   = \gamma ( Z^\mu   + v^\mu),\quad Z^\mu v_\mu=0,
\end{equation}
where 
\begin{equation}\label{key}
v^\mu   \equiv \frac{\Delta^\mu{}_\nu u^\nu}{\gamma} , \qquad 
\Delta_{\mu \nu}   \equiv g_{\mu \nu}   +\frac{1}{c^2} Z_\mu Z_\nu  .
\end{equation}
We may introduce a local orthonormal tetrad $\{e_{\hat{\mu}}\}$ \cite{Wald} at $P$ with 
$(e_{\hat{0}})^\mu = Z^\mu$, which implies the introduction of Cartesian coordinates in the 
(co)tangent space of the spacetime which is naturally endowed with a Minkowski metric.
Under this choice, $v^\mu$ can be expressed as $v^{\hat{\mu}}  =(0,\boldsymbol{v})$, and 
correspondingly, $u^{\hat{\mu}} = \gamma (c,\boldsymbol{v})$. It is evident that 
$v^\mu$ is the spatial velocity of the particle 
in coordinate time with respect to the observer at the event $P$. 
Using the normalization condition $u^{\hat{\mu}} u_{\hat{\mu}} = -c^2$, we have
\begin{equation}\label{gamma in v/c}
\gamma   = \frac{1}{\sqrt{1-\frac{v^2}{c^2}}}, \qquad v^2=\boldsymbol{v}\cdot \boldsymbol{v},
\end{equation}
where $\cdot$ means the Cartesian scalar product. It turns out that $\gamma$ is 
nothing but the local Lorentz factor in the tangent space of $\mathcal{M}$ at $P$. 
Naturally, $Z^\mu$ and $u^\mu$ are connected by a 
local Lorentz boost. 

\subsection{Power and work of the total external force} 

Let us start from the simple case of Minkowski spacetime. We assume that there 
is a nonvanishing external force $F^\mu$ exerted on the particle. The worldline 
of the particle is then non-geodesic and is described by the equation
\begin{equation}\label{F}
F^\mu = \frac{\rd{p}^\mu}{\rd{\tau}} = m u^\nu\nabla_\nu u^\mu,
\end{equation}
where $\tau$ is the proper time of the particle. Let us remind that in 
Minkowski spacetime, there is no difference between the covariant derivative 
$\nabla_\mu$ and the ordinary derivative $\partial_\mu$. We use solely the 
notation $\nabla_\mu$ in order that some of the results can be smoothly shifted 
to the cases of curved spacetimes.
	
For an inertial observer $\mathcal{O}_{\rm{i}}$ whose worldline intersects with the 
particle's worldline at the event $P$, we can extract the zeroth component of the 
force $F^\mu$ in Cartesian coordinates $x^\mu$ via
\[
F^0 = - z^\mu F_\mu ,
\]
where $z^\mu = (c,0,\cdots,0)$ is the proper velocity of the 
observer $\mathcal{O}_{\rm{i}}$. 
It is easy to recognize that $F^0$ is just the power of the 
external force $F^\mu$ defined as the proper time rate of the
particle's energy $E(\mathcal{O}_{\rm{i}})$ observed by the observer 
$\mathcal{O}_{\rm{i}}$,
\[
F^0 =  - z^\mu F_\mu  = \frac{\rd{E(\mathcal{O}_{\rm{i}})}}{\rd{\tau}},
\qquad
E(\mathcal{O}_{\rm{i}}) = - z^\mu p_\mu =-m z^\mu u_\mu .
\]

Now if we change the spacetime from Minkowskian to a generic curved manifold 
$\mathcal{M}$ and replace the inertial observer $\mathcal{O}_{\rm{i}}$ by a 
generic observer $\mathcal{O}$, the equation of motion of the particle is 
still described by \eqref{F}, but the combination $-Z^\mu F_\mu$ is no longer the 
zeroth component of $F^\mu$. In spite of this difference, it is still 
reasonable to define the power of the total external force at event $P$ as
\begin{equation}\label{rate Wex}
\mathcal{P}_{\rm{ex}}(\mathcal{O}) \equiv -Z^\mu F_\mu.
\end{equation} 
We can also make an orthogonal decomposition for $F^\mu$ with respect to $Z^\mu$, 
just like what we have done for $u^\mu$ in eq.\eqref{dec u},
\begin{equation}\label{dec F}
F^\mu  = -\frac{1}{c^2} \left(Z^\nu F_\nu\right) Z^\mu   + \gamma f^\mu  ,
\end{equation}
where 
\begin{equation}\label{key}
f^\mu   \equiv \frac{\Delta^\mu {}_\nu F^\nu}{\gamma} 
\end{equation}
is the spatial external force. Inserting eqs.\eqref{dec u} and \eqref{dec F} 
into eq.\eqref{rate Wex} yields
\begin{equation}\label{dWex=fv}
\mathcal{P}_{\rm{ex}}(\mathcal{O}) = \gamma f^\mu v_\mu.
\end{equation}
Using the tetrad $\{e_{\hat{\mu}}\}$, $f^\mu $ can be written as $(0,\boldsymbol{f})$, 
and eq.\eqref{dWex=fv} becomes
\begin{equation}\label{rate Wex cla}
\mathcal{P}_{\rm{ex}}(\mathcal{O}) = \gamma \boldsymbol{f}\cdot \boldsymbol{v}.
\end{equation}
This form of the power of the total external force in \eqref{rate Wex} is 
more conceivable due to the explicit resemblance to the corresponding Newtonian 
expression.

We have seen that, in order to define the local power of the total external force,
only a single observer $\mathcal{O}$ is needed. However, the situation is 
different while considering the work done by total external force, because the 
latter ought to be defined via integrating the local power along the 
particle's worldline. In order that the integration of the 
local power is well defined, a densely distributed set of observers needs to be 
introduced, and the worldline of each observer from this observer set must intersect 
the particle's worldline at a single event. Following \cite{Yu}, we define an observer 
field $\mathcal{Q}$ on $\mathcal{M}$ as a smooth timelike future directed vector 
field, each of whose integral curves is an observer. Form now on, $Z^\mu$ will
be used to denote the proper velocity field of $\mathcal{Q}$. With the above 
preparations, we can now write the work done by the total external force
during the proper time interval $[\tau_i,\tau_f]$ as
\begin{align}
W_{\rm{ex}}(\mathcal{Q}) 
=\int_{\tau_i}^{\tau_f} \mathcal{P}_{\rm ex}(\mathcal{Q}) \rd \tau
=-\int_{\tau_i}^{\tau_f} Z^\mu F_\mu\rd\tau.	
\end{align}

\subsection{Power and work of gravity via work-energy theorem}\label{def of work} 

As mentioned in the introduction, gravity is not viewed as a force in 
general relativity. Therefore, it looks hard to define the power of and the work 
done by gravity. To find a way out, let us recall that, in Newtonian mechanics, 
there is a well established law, i.e. the work-energy theorem, which states that
{\em the change in the energy of a particle equals the total work done on it}. 
In other words, the only way to change the energy of a particle is 
to exert mechanical work on it. This statement is among the very few basic 
principles on top of which classical mechanics is established. There is no reason 
why such a statement could get changed simply by shifting to 
relativistic systems. Therefore, we assume the work-energy theorem holds
and take it as the tool for defining the power of and the work done by gravity.

In relativistic context, the total work done on a particle is consisted of the work 
done by the total external force and that by gravity (or inertial force, thanks to 
Einstein's equivalence principle). Thus we have
\begin{align}
W_{\rm{grav}}(\mathcal{Q})+ W_{\rm{ex}}(\mathcal{Q}) 
= \Delta E(\mathcal{Q}), \label{w-e int}
\end{align}
where $W_{\rm{grav}}(\mathcal{Q})$ is the work done by gravity which is yet to 
be defined. In differential form, the work-energy theorem can be 
expressed as follows
\begin{equation}\label{w-e theorem}
\mathcal{P}_{\rm grav}(\mathcal{Q}) + \mathcal{P}_{\rm ex}(\mathcal{Q}) 
= \frac{\rd E(\mathcal{Q})}{\rd \tau},
\end{equation}
where $\mathcal{P}_{\rm grav}(\mathcal{Q})$ represent the local power of gravity
measured by the observer field $\mathcal{Q}$. Naturally, $W_{\rm{grav}}(\mathcal{Q})$
and $\mathcal{P}_{\rm grav}(\mathcal{Q})$ should be connected via
\begin{align}
W_{\rm{grav}}(\mathcal{Q})
=\int_{\tau_i}^{\tau_f} \mathcal{P}_{\rm grav}(\mathcal{Q}) \rd\tau.
\label{Wgrav}
\end{align}

Recall that $\mathcal{P}_{\rm ex}(\mathcal{Q}) $ is defined in 
eq.\eqref{rate Wex}. If the
proper time rate of $E(\mathcal{Q})$, i.e. $\rd E(\mathcal{Q})/\rd \tau$ could
be evaluated independently, then eq.\eqref{w-e theorem} would give rise to 
a definition for $\mathcal{P}_{\rm grav}(\mathcal{Q})$. Fortunately, 
$\rd E(\mathcal{Q})/\rd \tau$ can be evaluated right away from 
eq.\eqref{engy}, yielding
\begin{align}
\frac{\rd E(\mathcal{Q})}{\rd \tau}  
&=  - u^\nu \nabla_{\nu} \left( Z^\mu p_\mu\right)
= -m u_\mu \left(u^\nu \nabla_{\nu} Z^\mu\right)  
-m Z^\mu \left(u^\nu \nabla_\nu u_\mu\right) \nonumber \\
&= -mu_\mu \left(u^\nu \nabla_{\nu} Z^\mu\right) - Z^\mu F_\mu .
\label{change of kine}
\end{align}
The last term on the RHS of eq.\eqref{change of kine} is precisely equal to 
$\mathcal{P}_{\rm ex}(\mathcal{Q})$. Therefore, the first term on the RHS 
of eq.\eqref{change of kine} needs to be equal 
to $\mathcal{P}_{\rm grav}(\mathcal{Q})$
in order for the work-energy theorem to hold,
\begin{align}
\mathcal{P}_{\rm grav}(\mathcal{Q}) =-mu_\mu \left(u^\nu \nabla_{\nu} Z^\mu\right).
\label{Pgrav}
\end{align}
Inserting eq.\eqref{Pgrav} into \eqref{Wgrav}, we get
\begin{equation}\label{wgrav}
W_{\mathrm{grav}}(\mathcal{Q}) 
= \int_{\tau_i}^{\tau_f} \mathcal{P}_{\rm grav}(\mathcal{Q}) \,\rd{\tau} 
= -\int_{\tau_i}^{\tau_f} m u^\mu u^\nu \nabla_\nu Z_\mu \,\rd\tau.
\end{equation}
This completes the definition for the work done by gravity. 

Before finishing this subsection, let us stress that the work-energy theorem 
should be regarded as a law of {\em Nature} which does not depend on the choice 
of observer field. The inclusion of the notation $\mathcal{Q}$ in
eq.\eqref{w-e int} is simply intended for reminding the fact that the 
values of the quantities $W_{\rm{grav}}(\mathcal{Q})$, $W_{\rm{ex}}(\mathcal{Q})$
and $ \Delta E(\mathcal{Q})$ are all dependent on the observer field, however 
the identity \eqref{w-e int} holds for any $\mathcal{Q}$.

\section{Examples}

In order to justify the definitions made in the last section, we now 
consider the motion of a probe particle in several well-known 
spacetime solutions of general relativity and calculate the corresponding 
gravitational work as well as work done by external forces (when applicable). It will 
be clear that in the far field limit, the works calculated using our 
definitions agree with the well-known non-relativistic result. These 
example cases may serve as a justification to our definitions.

\subsection{1+1 dimensional Minkowski spacetime}
To begin with, let us study two simple cases in Minkowski spacetime in order 
to have a quick intuitive understanding about our previous results. 

The first case involves a particle of mass $m$ moving in Minkowski spacetime 
with a nonzero 
external force exerted on it. We use the Cartesian coordinates $x^\mu=(ct,x)$, 
and as the first and simplest example case, the observer field is taken to 
be inertial with $Z^\mu = (c,0)$. The proper velocity of the particle can be 
written as $u^\mu = \gamma c(1, \tanh(a \tau /c))$,
where $\gamma=\cosh(a\tau/c)$. Clearly, the external force takes the form
\[
F^\mu=ma^\mu=mu^\nu\nabla_\nu u^\mu= ma \left[\sinh \left(\frac{a \tau}{c}\right), 
\cosh \left(\frac{a \tau}{c}\right)\right].
\]
Using the decomposition \eqref{dec F}, we get 
the spatial part of the force 
\begin{equation}\label{key}
f^\mu = (0, ma).
\end{equation}
The constant $a$ is nothing but the magnitude of the proper acceleration of the particle,
\begin{equation}\label{key}
\sqrt{a^\mu a_\mu} = a.
\end{equation}
The inertial observers perceive neither gravity nor inertial forces, hence the 
only relevant power comes from the external force which reads
\begin{equation}\label{key}
-Z^\mu F_\mu = mac \sinh \left(\frac{a \tau}{c}\right).
\end{equation}
Consequently the work done by external force is
\begin{align}
W_{\mathrm{ex}}(\mathcal{Q}) 
&= -\int_{\tau_{i}}^{\tau_{f}} Z^\mu F_\mu \rd \tau
= mc^2 \left[\cosh \left(\frac{a \tau_{f}}{c}\right)
- \cosh \left(\frac{a \tau_{i}}{c}\right)\right].
\label{mink ex}
\end{align}
In the slow motion limit \textcolor{blue}{$a\tau/c\to 0$}, $\tau$ is approximately equal to the 
coordinate time $t$, and we have
\[
u^\mu \simeq (c, at),
\]
and $W_{\mathrm{ex}}(\mathcal{Q})$ becomes
\[
W_{\mathrm{ex}}(\mathcal{Q})\simeq \frac{1}{2}m a^2(t_f^2-t_i^2),
\]
wherein we recognize that the RHS is the change in the kinematic energy of the particle,
which is the correct Newtonian result.

In the second case, the particle is kept fixed at the origin, with its 
worldline parametrized as $x^\mu=(c\tau, 0)$, so that $u^\mu = (c,0)$, 
while the observer field $\mathcal{Q}$ is subject to nontrivial acceleration. 
The proper velocity field of $\mathcal{Q}$ is given by
$Z^\mu=\gamma c(1,-\tanh(a\tau/c))$, where $\gamma=\cosh(a\tau/c)$.
Due to the accelerated motion of $\mathcal{Q}$, the observers in $\mathcal{Q}$ 
should observe an effective gravity (which is actually the inertial force) 
exerted on the particle. The work done by this fictitious gravity can be 
easily evaluated using eq.\eqref{wgrav}, yielding
\begin{align}
W_{\mathrm{grav}}(\mathcal{Q}) &= 
-\int_{\tau_i}^{\tau_f} m u^\mu u^\nu \nabla_\nu Z_\mu \,\rd\tau
= mc^2 \left[\cosh \left(\frac{a \tau_{f}}{c}\right) 
- \cosh \left(\frac{a \tau_{i}}{c}\right)\right]. \label{mink grav}
\end{align}
Notice that in both cases the amount of works are the same, though they have 
completely different interpretations. We can of course take the slow motion limit 
in this case and verify that $\displaystyle W_{\mathrm{grav}}(\mathcal{Q})\simeq 
\frac{1}{2}m a^2(t_f^2-t_i^2)$ in this limit.

\begin{comment}	
It should be emphasized that even Minkowski spacetime is flat, 
there seems to be a nonzero ``gravitational work'', which is actually 
fictitious because it only owes to the noninertial movement of the observers. 
However, as will be shown in the next section, in a curved spacetime the work 
done by inertial force can not be distinguished from the real gravitational work. 
\end{comment}

\subsection{Schwarzschild spacetime}

The previous example is trivial in the sense that there is no actual
gravity involved in it. The fictitious gravity is solely triggered by the 
accelerated motion of the observer field. 
The next example is nontrivial because the system 
involves real gravity. 

Consider a probe particle of mass $m$ moving in Schwarzschild spacetime 
with line element
\begin{align}
\rd s^{2}=-\left(1-\frac{r_{g}}{r}\right) c^2\rd t^{2}
+\left(1-\frac{r_{g}}{r}\right)^{-1}\rd r^{2}+r^{2}\rd\Omega_{2},
\end{align}
where $r_g=2GM/c^2$. The Lagrangian of the 
probe particle reads
\begin{align}\label{free L for probe}
L=\frac{m}{2}g_{\mu\nu}\dot{x}^{\mu}\dot{x}^{\nu}
=\frac{m}{2}\left[-\left(1-\frac{r_{g}}{r}\right)c^2 \dot{t}^{2}
+\left(1-\frac{r_{g}}{r}\right)^{-1}\dot{r}^{2}
+r^{2}\left(\dot{\theta}^{2}+\sin^{2}\theta\dot{\phi}^{2}\right)\right],
\end{align} 
where the over dots represent derivatives with respect to the proper 
time $\tau$. This Lagrangian describes a ``free'' (i.e. in the absence 
of external forces besides gravity) probe particle, whose worldline 
follows one of the geodesics. The equations of motion can be 
obtained by straightforward variations of eq.\eqref{free L for probe}. 
However, it is better to consider the 
first integrals and the on-shell condition as alternatives for the 
Euler-Lagrangian equations. 

Without loss of generality, we assume that the probe particle 
moves in the equatorial plane $\theta=\pi/2$. Then the Lagrangian
is reduced into
\begin{align}\label{key}
L=\frac{m}{2}\left[-\left(1-\frac{r_{g}}{r}\right)c^2 \dot{t}^{2}
+\left(1-\frac{r_{g}}{r}\right)^{-1}\dot{r}^{2}+r^2\dot{\phi}^2\right].
\end{align}
Since both $t$ and $\phi$ are Killing coordinates, we have the
corresponding integrals of motion
\begin{align}
&\frac{\partial L}{\partial\dot{t}}
=-m\left(1-\frac{r_{g}}{r}\right)c^2 \dot{t}
=-\mathcal{E}, \label{pL1}\\
&\frac{\partial L}{\partial\dot{\phi}}=mr^2\dot{\phi}=\mathcal{J}.
\label{pL2}
\end{align}
The constants $\mathcal{E,J}$ respectively can be understood as the 
energy and the angular momentum ``measured'' by the Killing vector 
fields $k=\partial_t$ and $\chi= \partial_\phi$, and 
their values can only be settled by initial conditions. 

The on-shell condition is expressed as follows:
\begin{align}\label{Schw on-shell}
u^{\mu}u_{\mu}=-\left(1-\frac{r_{g}}{r}\right)c^2\dot{t}^{2}
+\left(1-\frac{r_{g}}{r}\right)^{-1}\dot{r}^{2} + r^2 \dot{\phi}^2=-c^2.
\end{align}
Using eqs.\eqref{pL1}, \eqref{pL2} and \eqref{Schw on-shell}, 
we get
\begin{align}
&\dot{t} = \frac{\mathcal{E}}{mc^2}\left(1-\frac{r_{g}}{r}\right)^{-1},
\label{tdot}\\
&\dot{r}=-\sqrt{\left(\frac{\mathcal{E}}{mc}\right)^{2}
-\left(1-\frac{r_g}{r}\right)\left(c^2+\frac{\mathcal{J}^2}{m^2r^2}\right)},
\label{rdot}
\end{align}
where the total minus sign on the RHS of eq.\eqref{rdot}
indicates that the particle falls inwardly. 

We assume the following initial 
conditions
\begin{align*}
&x^\mu(\tau_i) = \left(ct_i, r_i, \frac{\pi}{2},\phi_i\right),\\
&u^\mu(\tau_i)=\left[c\left(1-\frac{r_{g}}{r_i}\right)^{-1/2},
0,0,0\right],
\end{align*}
where $t_i,\phi_i$ are arbitrary constants whose values are taken
within the allowed ranges for the respective coordinates and 
meanwhile the constant $r_i>r_g$. The above choice for 
$u^\mu(\tau_i)$ is compatible with the on-shell condition.
Using the above initial conditions and recalling the constancy of 
$\mathcal{E,J}$, we can determine that 
\begin{align}
\mathcal{E}=mc^2\left(1-\frac{r_{g}}{r_i}\right)^{1/2},
\qquad
\mathcal{J}=0.
\label{igm}
\end{align}

In order to examplify that the power and work can be defined and evaluated for 
generic observer fields rather than just for the static observer field, let us take
the observer field $\mathcal{Q}$ to be moving outward, with proper velocity
\begin{equation}\label{ob}
Z^\mu =c \left[ \left(1-\frac{r_g}{r}\right)^{-1} 
\left(1+C-\frac{r_g}{r}\right)^{1/2}, \sqrt{C}, 0, 0\right], \qquad (r>r_g)
\end{equation}
where $C>0$ is a dimensionless constant.

The power of the gravity can be evaluated straightforwardly using eq.\eqref{Pgrav},
which results in
\begin{align}
\mathcal{P}_{\rm grav}(\mathcal Q)
&=
\frac{m r_g c^3}{2(r-r_g)^2 \left(1+C-\frac{r_g}{r}\right)^{1/2}} \Bigg[\sqrt{C} \left(1+C-\frac{r_g}{r}\right)^{1/2} \left(\frac{2 \mathcal{E}^2}{m^2c^4}-1+\frac{r_g}{r}\right) \nonumber \\
&\qquad + \frac{\mathcal{E}}{mc^2}  \left( \frac{\mathcal{E}^2 }{m^2c^4}-1+\frac{r_g}{r}\right)^{1/2} \left(2C+1 -\frac{r_g}{r}\right) \Bigg].
\end{align}
The corresponding work is
\begin{align}
W_{\mathrm{grav}}(\mathcal{Q}) 
&= \int_{\tau_{i}}^{\tau_{f}} \mathcal{P}_{\rm grav}(\mathcal Q) \rd\tau \nonumber\\
%&= -\int_{r_{i}}^{r_{f}} \frac{m r_g c^2 \left(1+C-\frac{r_g}{r}\right)^{-1/2}}{2(r-r_g)^2} \Bigg[ \sqrt{C} \left(1+C-\frac{r_g}{r}\right)^{1/2} \left(\frac{\mathcal{E}^2}{m^2c^4}-1+\frac{r_g}{r}\right)^{-1/2} \nonumber \\
%&\qquad\times \left(\frac{2 \mathcal{E}^2}{m^2c^4} -1 +\frac{r_g}{r}\right) + \frac{\mathcal{E}}{mc^2} \left(2C+1 -\frac{r_g}{r}\right)\Bigg] d r\nonumber\\
&= mc^2 \left(1-\frac{r_g}{r}\right)^{-1} \left[\frac{\mathcal{E}}{mc^2} \left(1+C-\frac{r_g}{r}\right)^{1/2} + \sqrt{C} \left( \frac{\mathcal{E}^2}{m^2c^4} -1 +\frac{r_g}{r}\right)^{1/2}\right] \Bigg|_{r_i} ^{r_f}, \label{W1}
\end{align}
where $r_f =r(\tau_f)>r_g$ in order to avoid the
horizon. 
\begin{comment}
Substituting the initial condition \eqref{igm} into eq.\eqref{W1}, we have
\begin{align}
W_{\mathrm{grav}}(\mathcal{Q})
&= mc^2 \left(1-\frac{r_g}{r_i}\right)^{1/2} \left(1-\frac{r_g}{r}\right)^{-1} \left(1+C-\frac{r_g}{r}\right)^{1/2} \Bigg|_{r_i} ^{r_f} \nonumber\\
&\qquad+ mc^2 \sqrt{C}\left(1-\frac{r_g}{r_f}\right)^{-1} \left(\frac{r_g}{r_f}- \frac{r_g}{r_i}\right)^{1/2} . \label{schw wg}
\end{align}	
\end{comment}

In the far field limit $r_i>r_f \gg r_g$, the work $W_{\mathrm{grav}}(\mathcal{Q})$ 
can be expand as Taylor series
in $r_g/r_f $ and $r_g/r_i$. At the order $O(r_g/r_{i,f})$, we have
\begin{align}\label{ffW}
W_{\mathrm{grav}}(\mathcal{Q})
=mc^2\left[\frac{2C+1}{\sqrt{C+1}} \left(\frac{r_g}{r_f}- \frac{r_g}{r_i}\right) 
+ \sqrt{C}\sqrt{\frac{r_g}{r_f}-\frac{r_g}{r_i}}\,\right],
\end{align}
where we have used eq.\eqref{igm} to express $\mathcal{E}$ in terms of $r_g/r_i$.

Alternatively, if we take the limit $C\to 0$, $\mathcal{Q}$ degenerates into the
static observer field, whose proper velocity field is proportional to the 
timelike Killing vector field  $k=\partial_t$, i.e. 
\begin{equation}\label{key}
\lim\limits_{C\rightarrow 0} Z^\mu=[c(-g_{tt})^{-1/2},0,0,0]
=\left[c\left(1-\frac{r_g}{r}\right)^{-1/2},0,0,0\right].
\end{equation}
In this case, the gravitational work becomes
\begin{align}
\lim\limits_{C\rightarrow 0}  W_{\mathrm{grav}}(\mathcal{Q}) = \mathcal{E}\left[\left(1-\frac{r_{g}}{r_f }\right)^{-1/2}-\left(1-\frac{r_{g}}{r_i}\right)^{-1/2}\right]. \label{Wg C=0}
\end{align}
Notice that this is the exact result for gravitational work measured by 
static observer field. Now if we further take the far field 
limit $r_i>r_f \gg r_g$, the work becomes
\begin{align}\label{key}
W_{\mathrm{grav}}(\mathcal{Q})
\simeq GMm\left(\frac{1}{r_f }-\frac{1}{r_i}\right).
\end{align}
This result agrees with eq.\eqref{Ng} which describes the work 
done by Newtonian gravity. The same result will arise if we directly set $C\to 0$ in 
the far field limit result \eqref{ffW}.

Notice that, in eqs.\eqref{W1} and \eqref{Wg C=0}, there is an explicit dependence 
on $\mathcal{E}$, i.e. on the initial conditions. This dependence
seems to be indicating that the gravitational work in 
Schwarzschild spacetime is path dependent. Since 
Newtonian gravity is understood as a conservative force, 
the path dependence of gravitational work in the context of 
general relativity needs some explanations. 

Let us recall that, in Newtonian mechanics, the mechanical energy 
is divided into two parts, i.e. the kinematic and the potential 
energies. However, there is no corresponding description in 
general relativity. For simplicity, let us temporarily
assume $C=0$ and $\dot{\phi}=0$. 
Then, using eqs.\eqref{engy} and \eqref{tdot}, we get
\begin{align}\label{E in Schw}
E(\mathcal{Q})&=mc^2\dot{t}\left(1-\frac{r_g}{r}\right)^{1/2}
=\mathcal{E}\left(1-\frac{r_g}{r}\right)^{-1/2} \nonumber \\
&=mc^2\left[1-\frac{\dot{r}^2}{c^2\dot{t}^2
\left(1-\frac{r_g}{r}\right)^2}\right]^{-1/2}
=\frac{mc^2}{\sqrt{1-\frac{|\boldsymbol{v}|^2}{c(r)^2}}},
\end{align}
where $\displaystyle|\boldsymbol{v}|=\frac{\dot{r}}{\dot{t}}
$ is the spatial coordinate speed of the particle, and
\begin{align*}
c(r)=c\left|1-\frac{r_g}{r}\right|
\end{align*}
is the coordinate speed of light. 
Obviously eq.\eqref{E in Schw} has the form of \eqref{E=gam}, 
with the local Lorentz factor 
$\gamma= \frac{1}{\sqrt{1-\frac{|\boldsymbol{v}|^2}{c(r)^2}}}$.
According to eq.\eqref{E in Schw}, the initial energy of the particle 
can be written as
\begin{align*}
E(\mathcal{Q})\Big|_{\mathrm{init}} =\gamma_{\rm init} mc^2 
=\frac{mc^2}{\sqrt{1-\frac{|\boldsymbol{v}(\tau_i)|^2}
{c^2\left(1-\frac{r_g}{r_i}\right)^2}}}.
\end{align*} 
The initial Lorentz factor $\gamma_{\rm init}$ is clearly dependent on the 
initial condition. Since gravity acts on the physical 
mass rather than on the rest mass, it is reasonable that the work 
also relies on the initial condition. 
In fact, the mass of the particle changes as it moves towards the source, 
the effect of gravitational work also changes. 
If the energy of the particle becomes large enough, it 
should no longer be regarded as a probe, as its energy could make 
significant changes to the spacetime metric. 
This makes an important difference between general relativity and 
Newtonian mechanics. To see this more clearly, 
we assume $|\boldsymbol{v}(\tau_i)|^2 \neq 0$, so the initial energy is larger 
than $mc^2$ but not too large to break the probe approximation, 
the gravitational work done on such a particle can be evaluated to be
\begin{align}\label{wg for Schw,u(0) not 0}
W_{\mathrm{grav}}(\mathcal{Q})
=\frac{mc^2}{\sqrt{1-\frac{|\boldsymbol{v}(\tau_i)|^2}
{c^2\left(1-\frac{r_g}{r_i}\right)^2}}}
\left[\left(1-\frac{r_{g}}{r_i}\right)^{1/2}
\left(1-\frac{r_{g}}{r_f }\right)^{-1/2}-1\right].
\end{align}
Now we assume that there is another particle with rest mass $m'$, 
\begin{align*}
m'=\frac{m}{\sqrt{1-\frac{|\boldsymbol{v}(\tau_i)|^2}
{c^2\left(1-\frac{r_g}{r_i}\right)^2}}},
\end{align*}
which carries on a ``freely falling'' procedure, 
beginning with zero spatial velocity
at $r=r_i$. In this case, the initial energy of the second probe particle 
is just $m'c^2$. Then the work of gravity is
\begin{align}\label{wg rest,Schw}
W'_{\mathrm{grav}}(\mathcal{Q})
&=- \int_{\tau_i}^{\tau_f} m' u'_\mu u'^\nu\nabla_\nu Z^\mu \rd{\tau'}
\nonumber\\
&=\frac{mc^2}{\sqrt{1-\frac{|\boldsymbol{v}(\tau_i)|^2}
{c^2\left(1-\frac{r_g}{r_i}\right)^2}}}
\left[\left(1-\frac{r_{g}}{r_i}\right)^{1/2}
\left(1-\frac{r_{g}}{r_f }\right)^{-1/2}-1\right],
\end{align}
which is identical to \eqref{wg for Schw,u(0) not 0}. 
Therefore, the gravitational work exerted on a probe particle with 
nonzero initial velocity can be equivalently regarded as work exerted on 
a particle with zero initial velocity with rest mass equal to the 
physical mass of the former one.  
If the initial condition changes, the gravitational work will 
also change.

\subsection{Reisner-Nordstr\"om (RN) spacetime}

Now let us consider the \textcolor{blue}{third} example, i.e. 
a particle with mass $m$ and charge $e$ moving in RN
spacetime. The metric is given by 
\begin{align}\label{metric-RN}
\rd s^{2}=-\left(1-\frac{r_{g}}{r}+\frac{r_{Q}^{2}}{r^{2}}\right)
c^2\rd t^{2}
+\left(1-\frac{r_{g}}{r}+\frac{r_{Q}^{2}}{r^{2}}\right)^{-1}
\rd r^{2}+r^{2}\rd\Omega_{2},
\end{align}
which is accompanied by the electric potential
\begin{align}\label{key}
A_{\mu}=\left(\frac{Q}{4\pi \varepsilon_0 c r},0,0,0\right),
\end{align}
where $r_g$ and $r_Q$ are given respectively by
\begin{align}\label{rQ}
r_g=\frac{2GM}{c^2},\qquad
r_{Q}^{2}=\frac{GQ^{2}}{4\pi\varepsilon_0 c^4}.
\end{align}
The Lagrangian of a charged probe particle can be written as
\begin{align}\label{Lag for free+EM}
L\left(x,\dot{x}\right)=\frac{m}{2}g_{\mu\nu}\dot{x}^{\mu}\dot{x}^{\nu}
-eA_{\mu}\dot{x}^{\mu},
\end{align}
and the corresponding equation of motion is given by
\begin{align}\label{key}
mu^{\nu}\nabla_{\nu}u^{\mu}=-eF^{\mu\nu}u_{\nu}\equiv F^\mu,
\end{align}
where $F_{\mu\nu}=\nabla_{\mu}A_{\nu}-\nabla_{\nu}A_{\mu}$ is 
the field strength tensor and $F^\mu$ is the total electromagnetic force 
acting on the charged probe particle. 
Since the background spacetime is spherically symmetric, we can still
assume that the probe particle moves in the equatorial plane 
$\theta=\pi/2$, $\dot{\theta}=0$. Then in explicit form the
Lagrangian can be re-written as
\begin{align}
L\left(x,\dot{x}\right)
=\frac{m}{2}\left[-\left(1-\frac{r_{g}}{r}
+\frac{r_{Q}^{2}}{r^{2}}\right)c^2\dot{t}^{2}
+\left(1-\frac{r_{g}}{r}+\frac{r_{Q}^{2}}{r^{2}}\right)^{-1}\dot{r}^{2}
+r^2\dot{\phi}^2\right]-\frac{eQ}{4\pi\varepsilon_0 r}\dot{t}.
\end{align} 
The on-shell condition now reads
\begin{align}\label{on-shell for RN}
u^{\mu}u_{\mu}=-\left(1-\frac{r_{g}}{r}
+\frac{r_{Q}^{2}}{r^{2}}\right)c^2\dot{t}^{2}+\left(1-\frac{r_{g}}{r}
+\frac{r_{Q}^{2}}{r^{2}}\right)^{-1}\dot{r}^{2}+r^2\dot{\phi}^2=-c^2.
\end{align}
The integrals of motion associated with 
the Killing coordinates $t,\phi$ are given by
\begin{align}\label{E1 def}
&\frac{\partial L}{\partial\dot{t}}=-m\left(1-\frac{r_{g}}{r}
+\frac{r_{Q}^{2}}{r^{2}}\right)c^2\dot{t}-\frac{eQ}{4\pi\varepsilon_0 r}
\equiv -\mathcal{E}_{1},\\
\label{J1 def}
&\frac{\partial L}{\partial\dot{\phi}}
=mr^2\dot{\phi}\equiv \mathcal{J}_{1},
\end{align}
and from the above equations we can get
\begin{align}\label{t dot for RN}
\dot{t}=\frac{\left(\mathcal{E}_{1}
-\frac{eQ}{4\pi\varepsilon_0 r}\right)}{mc^2}\left(1-\frac{r_{g}}{r}
+\frac{r_{Q}^{2}}{r^{2}}\right)^{-1},\qquad
\dot{\phi}=\frac{\mathcal{J}_{1}}{mr^2}.
\end{align}
Using \eqref{t dot for RN} and \eqref{on-shell for RN} we get
\begin{align}\label{r dot for RN}
\dot{r}=-\left[\frac{\left(\mathcal{E}_{1}
-\frac{eQ}{4\pi\varepsilon_0 r}\right)^{2}}{m^{2}c^2}
-\left(1-\frac{r_{g}}{r}+\frac{r_{Q}^{2}}{r^{2}}\right)
\left(c^2+\frac{\mathcal{J}_{1}^2}{m^2r^2}\right)\right]^{1/2}.
\end{align}
The initial conditions are now chosen as
\begin{align}
&x^\mu(\tau_i) = \left(c t_i, r_i, \frac{\pi}{2},\phi_i\right),\\
&u^\mu(\tau_i)=\left[c\left(1-\frac{r_{g}}{r_i}
+\frac{r_{Q}^{2}}{r_i^{2}}\right)^{-1/2},
0,0,0\right], \label{initRN}
\end{align}
where again $t_i,\phi_i$ take arbitrary allowed constant values 
within the ranges for the respective coordinates and 
$r_i>r_+$ ($r_+$ is the radius of the event horizon of RN black hole).
The proper velocity field of $\mathcal{Q}$ is chosen to be 
\begin{equation}\label{key}
Z^\mu = c \left[\left(1-\frac{r_{g}}{r} +\frac{r_{Q}^{2}}{r^{2}}\right)^{-1}\left(1+C_1 -\frac{r_g}{r} +\frac{r_Q^2}{r^2}\right)^{1/2},\sqrt{C_1},0,0\right],\qquad (r>r_+)
\end{equation}
where $C_1>0$ is a dimensionless constant. Therefore, 
the powers of the electromagnetic force and that of the 
gravity are given respectively as
\begin{align*}
\mathcal{P}_{\rm em}(\mathcal Q)
&= -\frac{eQc}{4\pi \varepsilon_0 \left(r^2- r_g r+ r_Q^2\right)} \Bigg[ \frac{\sqrt{C_1}}{mc^2} \left(\mathcal{E}_1 -\frac{eQ}{4\pi\varepsilon_0 r}\right) \\
&\qquad+ \left(1+C_1-\frac{r_g}{r}+\frac{r_Q^2}{r^2}\right)^{1/2} \Bigg(\frac{\left(\mathcal{E}_1 -\frac{eQ}{4\pi\varepsilon_0 r}\right)^2}{m^2c^4} - 1+\frac{r_g}{r} - \frac{r_Q^2}{r^2}\Bigg)^{1/2}\Bigg],\\
\mathcal{P}_{\rm grav}(\mathcal Q)
&= \frac{mc^3r \left(r_g r-2r_Q^2\right)}{2 \left(r^2-r_g r +r_Q^2\right)^2 \left(1+C_1-\frac{r_g}{r}+\frac{r_Q^2}{r^2}\right)^{1/2}} \Bigg[\frac{\mathcal{E}_1 - \frac{eQ}{4\pi\varepsilon_0 r}}{mc^2} \left(2C_1 +1 -\frac{r_g}{r}+\frac{r_Q^2}{r^2}\right) \\
 &\qquad\times \Bigg(\frac{\left(\mathcal{E}_1 -\frac{eQ}{4\pi\varepsilon_0 r}\right)^2}{m^2c^4} - 1+\frac{r_g}{r} - \frac{r_Q^2}{r^2}\Bigg)^{1/2} \\
 &\qquad+ \sqrt{C_1}\left(1+C_1-\frac{r_g}{r}+\frac{r_Q^2}{r^2}\right)^{1/2} 
 \Bigg(\frac{2 \left(\mathcal{E}_1 -\frac{eQ}{4\pi\varepsilon_0 r}\right)^2}{m^2c^4} - 1+\frac{r_g}{r} - \frac{r_Q^2}{r^2}\Bigg)\Bigg].
\end{align*}
The corresponding works are given by the integrals of the above powers over 
the proper time interval $[\tau_i,\tau_f]$, however these integrals cannot 
be worked out explicitly due to the overwhelming complexities of the integrands. 

The situation will get simplified drastically in the following two special cases. 

The first special case is when the probe particle is neutral, i.e. $e=0$.
In this case, $\mathcal{P}_{\rm em}(\mathcal{Q})=0$ and 
hence $W_{\mathrm{em}}(\mathcal{Q})=0$, and $W_{\mathrm{grav}}(\mathcal{Q})$ can be 
evaluated explicitly, 
\begin{align}
W_{\mathrm{grav}}(\mathcal{Q})
=&\left(1-\frac{r_g}{r}+\frac{r_Q^2}{r^2}\right)^{-1} 
\Bigg[\mathcal{E}_1 \left(1+ C_1-\frac{r_g}{r}+\frac{r_Q^2}{r^2}\right)^{1/2} 
\nonumber\\
&+\sqrt{C_1}mc^2 \left(\frac{ \mathcal{E}_1^2}{m^2c^4}-1+\frac{r_g}{r}
-\frac{r_Q^2}{r^2}\right)^{1/2}\Bigg] \Bigg|_{r_i}^{r_f}.
\label{RNgwork}
\end{align}
The gravitational work still contains contribution from the charge $Q$ because 
$Q$ not only produces electromagnetic field but also contributes to the gravity of
the background spacetime.
It remains to determine the integral of motion $\mathcal{E}_1$ 
using the initial data \eqref{initRN}. Using eqs.\eqref{initRN} and
\eqref{t dot for RN}, we get
\begin{align}\label{E1}
\mathcal{E}_{1}=mc^2\left(1-\frac{r_{g}}{r_i}
+\frac{r_{Q}^{2}}{r_i^{2}}\right)^{1/2}+\frac{eQ}{4\pi\varepsilon_0 r_i}.
\end{align}
Inserting this result into eq.\eqref{RNgwork} would finish the 
evaluation of the work done by gravity in RN spacetime.
If $Q=0$, then eq.\eqref{RNgwork} will fall back to the corresponding result
\eqref{W1} in the Schwarzschild case if we identify $C_1$ with $C$.
\begin{comment}
The second special case is the far field limit $r_i > r_f \gg r_g>2r_Q$. 
In this case, at the leading order $O(r_g/r_{i,f})$ and $O(r_Q/r_{i,f})$, we have
\begin{align}
W_{\mathrm{em}}(\mathcal{Q})&= -\frac{eQ}{4\pi\varepsilon_0}\left[\left(\frac{eQ}{2\pi\varepsilon_0 mc^2}-r_g\right)^{-1/2}\sqrt{C_1 r_i} +\sqrt{C_1+1}\right]\left(\frac{1}{r_f}-\frac{1}{r_i}\right),\\
W_{\mathrm{grav}}(\mathcal{Q})&= G M m \left[\frac{2 C_1 +1}{\sqrt{C_1 +1}}+ \left(\frac{eQ}{2\pi\varepsilon_0 mc^2}-r_g\right)^{-1/2}\sqrt{C_1 r_i}\right]\left(\frac{1}{r_f}-\frac{1}{r_i}\right).
\end{align}
\end{comment}

The second special case is to take the static observer field from the very beginning,
i.e. choosing $C_1=0$ and 
\begin{align}\label{static observer for RN}
Z^{\mu}=\left[c\left(1-\frac{r_{g}}{r}
+\frac{r_{Q}^{2}}{r^{2}}\right)^{-1/2},0,0,0\right].
\end{align}
Then we will have
\begin{align}
W_{\mathrm{em}}(\mathcal{Q})&= \int_{r_i}^{r_f }\frac{eQ}{4\pi\varepsilon_0 r^{2}}
\left(1-\frac{r_{g}}{r}+\frac{r_{Q}^{2}}{r^{2}}\right)^{-1/2}
\rd r\nonumber\\
&=\frac{eQ}{8\pi\varepsilon_0 r_{Q}}\ln\left(\frac{2r_{Q}^{2}-r_{g}r
-2r_{Q}\sqrt{r^{2}-r_{g}r+r_{Q}^{2}}}{2r_{Q}^{2}-r_{g}r
+2r_{Q}\sqrt{r^{2}-r_{g}r+r_{Q}^{2}}}\right)
\Bigg|_{r_i}^{r_f },
\label{Wex for RN C=0}
\end{align}
and
\begin{align}
W_{\mathrm{grav}}(\mathcal{Q})
&=  \int_{r_i}^{r_f }-\frac{\left(4\pi\mathcal{E}_{1}\varepsilon_0 r-eQ\right)
\left(r_{g}r-2r_{Q}^{2}\right)}{8\pi\varepsilon_0 r^{4}}\left(1-\frac{r_{g}}{r}
+\frac{r_{Q}^{2}}{r^{2}}\right)^{-3/2}\rd r \nonumber\\
&= \left[\mathcal{E}_{1}-\frac{eQ}{4\pi\varepsilon_0}\left(\frac{1}{r}
-\frac{1}{r_i}\right)\right]\left(1-\frac{r_g}{r}
+\frac{r_Q^2}{r^2}\right)^{-1/2}\Bigg|_{r_i}^{r_f }\nonumber\\
&\qquad-\frac{eQ}{8\pi\varepsilon_0 r_{Q}}\ln\left(\frac{2r_{Q}^{2}-r_{g}r
-2r_{Q}\sqrt{r^{2}-r_{g}r+r_{Q}^{2}}}{2r_{Q}^{2}-r_{g}r
+2r_{Q}\sqrt{r^{2}-r_{g}r+r_{Q}^{2}}}\right)\Bigg|_{r_i}^{r_f }.
\label{Wg for RN C=0}
\end{align}
Notice that the last term of eq.\eqref{Wg for RN C=0} is equal to
$-W_{\mathrm{em}}(\mathcal{Q})$. Therefore, 
the total work becomes
\begin{align*}
\Delta E(\mathcal{Q})
=&\left[\mathcal{E}_{1}-\frac{eQ}{4\pi\varepsilon_0}\left(\frac{1}{r}
-\frac{1}{r_i}\right)\right]\left(1-\frac{r_g}{r}
+\frac{r_Q^2}{r^2}\right)^{-1/2}\Bigg|_{r_i}^{r_f }.
\end{align*}
If the probe particle is neutral, $e=0$, then
\begin{align*}
W_{\mathrm{grav}}(\mathcal{Q})
=\mathcal{E}_{1}\left[\left(1-\frac{r_{g}}{r_f }
+\frac{r_{Q}^{2}}{r_f ^{2}}\right)^{-1/2}
-\left(1-\frac{r_{g}}{r_i}
+\frac{r_{Q}^{2}}{r_i^{2}}\right)^{-1/2}\right].
\end{align*}
If, in addition, $Q=0$, then
\begin{align*}
W_{\mathrm{grav}}(\mathcal{Q})
=\mathcal{E}_{1}\left[\left(1-\frac{r_{g}}{r_f }\right)^{-1/2}
-\left(1-\frac{r_{g}}{r_i}\right)^{-1/2}\right],
\end{align*}
which reproduces the result \eqref{Wg C=0} in the Schwarzschild case if we set 
$\mathcal{E}_1=\mathcal{E}$.
Taking expansion of eq.\eqref{Wex for RN C=0} and keeping the leading order yields
\begin{align}\label{Wex RN expand}
W_{\mathrm{em}}(\mathcal{Q}) \simeq
-\frac{eQ}{4\pi\varepsilon_0}\left(\frac{1}{r_f }-\frac{1}{r_i}\right),
\end{align}
This approximate result recovers the work done by the static electric field in 
Newtonian case.

\subsection{Kerr-Newman spacetime}

As the last example case, let us consider 
a probe particle with mass $m$ and charge $e$ moving in Kerr-Newman 
spacetime. The Lagrangian of the particle can be still written in the form 
\eqref{Lag for free+EM}, but with different underlying
spacetime geometry and different electromagnetic potential. 
In Boyer-Linquist coordinates, the metric reads:
\begin{align}\label{metric for Kerr-Newman}
{\rm d}s^{2}=-\frac{\Delta}{\rho^{2}}\left(c{\rm d}t
-a\sin^{2}\theta{\rm d}\phi\right)^{2}
+\frac{\sin^{2}\theta}{\rho^{2}}\left(\left(r^{2}+a^{2}\right){\rm d}\phi
-ac{\rm d}t\right)^{2}+\rho^{2}\left(\frac{{\rm d}r^{2}}{\Delta}
+{\rm d}\theta^{2}\right),
\end{align}
where
\begin{align}
&\rho^{2}=r^{2}+a^{2}\cos^{2}\theta,\\
&\Delta=r^{2}-r_{g}r+r_{Q}^{2}+a^{2},
\end{align}
and the parameters $r_g,r_Q$ are respectively connected with the 
mass $M$ and the charge $Q$ of the gravitational source and $a$ is
the ratio of the angular momentum $J$ over the mass $M$ of the source,
\begin{align}
a=\frac{J}{Mc}.	
\end{align}
In the above coordinate system, the electromagnetic potential of the 
source is given by
\begin{align}\label{A_mu for Kerr-Newman}
A_{\mu}=\left(\frac{Qr}{4\pi\varepsilon_0 c\rho^{2}},0,0,
-\frac{aQr\sin^{2}\theta}{4\pi\varepsilon_0 c\rho^{2}}\right).
\end{align}
Substituting eqs.\eqref{metric for Kerr-Newman}, \eqref{A_mu for Kerr-Newman} 
into eq.\eqref{Lag for free+EM}, we get the explicit form for the Lagrangian,
\begin{align}
L=&-\frac{m}{2}\left(1-\frac{r_{g}r-r_{Q}^{2}}{\rho^{2}}\right)c^2\dot{t}^{2}
+\frac{ma}{\rho^{2}}\sin^{2}\theta\left(-r_{g}r+r_{Q}^{2}\right)
c\dot{t}\dot{\phi}\nonumber\\
&+\frac{m}{2}\sin^{2}\theta\left(r^{2}+a^{2}
+\frac{a^{2}}{\rho^{2}}\sin^{2}\theta\left(r_{g}r
-r_{Q}^{2}\right)\right)\dot{\phi}^{2}
+\frac{m}{2}\rho^{2}\left(\frac{\dot{r}^{2}}{\Delta}+\dot{\theta}^{2}\right)
\nonumber \\
&-\frac{eQr}{4\pi\varepsilon_0 c\rho^{2}}\left(c\dot{t}-a\sin^{2}\theta\dot{\phi}\right).
\end{align}
This Lagrangian depends on neither $t$ nor $\phi$, therefore we have 
two integrals of motion:
\begin{align}
\frac{\partial L}{\partial\dot{t}}=&-m\left(1-\frac{r_{g}r
-r_{Q}^{2}}{\rho^{2}}\right)c^2\dot{t}+\frac{ma}{\rho^{2}}\sin^{2}\theta
\left(-r_{g}r+r_{Q}^{2}\right)c\dot{\phi}-\frac{eQr}{4\pi\varepsilon_0\rho^{2}}
\equiv -\mathcal{E}_{2},\label{E2}\\
\frac{\partial L}{\partial\dot{\phi}}
=&\, m\sin^{2}\theta\left(r^{2}+a^{2}
+\frac{a^{2}}{\rho^{2}}\sin^{2}\theta\left(r_{g}r
-r_{Q}^{2}\right)\right)\dot{\phi}\nonumber\\
&+\frac{ma}{\rho^{2}}\sin^{2}\theta\left(-r_{g}r
+r_{Q}^{2}\right)c\dot{t}
+\frac{eQra\sin^{2}\theta}{4\pi\varepsilon_0 c\rho^{2}}
\equiv\mathcal{J}_{2}. \label{J2}
\end{align}
It would be convenient if we restrict $\theta$ to $\pi/2$ so that 
$\rho=r$. Then we can solve $\dot{t}$ and $\dot{\phi}$ form eqs.\eqref{E2} 
and \eqref{J2}, yielding
\begin{align}\label{t dot for Kerr-Newman, theta=pi/2}
&\dot{t}=\frac{\frac{\mathcal{E}_{2}}{mc^2}\left[r^{2}
+a^{2}\left(1+\frac{r_{g}}{r}+\frac{r_{Q}^{2}}{r^{2}}\right)\right]
-\frac{eQ}{4\pi \varepsilon_0 rmc^2}\left(r^{2}+a^{2}
+\frac{2a^{2}r_{Q}^{2}}{r^{2}}\right)
+\frac{a\mathcal{J}_{2}}{mc}\left(-\frac{r_{g}}{r}
+\frac{r_{Q}^{2}}{r^{2}}\right)}{\Delta+\frac{2a^{2}r_{Q}^{2}}{r^{2}}
\left(1-\frac{r_{g}}{r}+\frac{r_{Q}^{2}}{r^{2}}\right)},\\
\label{phi dot for Kerr-Newman, theta=pi/2}
&\dot{\phi}=\frac{\frac{\mathcal{J}_{2}}{m}\left(1-\frac{r_{g}}{r}
+\frac{r_{Q}^{2}}{r^{2}}\right)-\frac{a\mathcal{E}_{2}}{mc}
\left(-\frac{r_{g}}{r}+\frac{r_{Q}^{2}}{r^{2}}\right)
-\frac{aeQ}{4\pi \varepsilon_0 rmc}}{\Delta+\frac{2a^{2}r_{Q}^{2}}{r^{2}}
\left(1-\frac{r_{g}}{r}+\frac{r_{Q}^{2}}{r^{2}}\right)}.
\end{align}
By applying the on-shell condition
\begin{align}
u_{\mu}u^{\mu}=&-\left(1-\frac{r_{g}}{r}+\frac{r_{Q}^{2}}{r^{2}}\right)c^2
\dot{t}^{2}+\frac{r^{2}}{\Delta}\dot{r}^{2}+\left[r^{2}
+a^{2}\left(1+\frac{r_{g}}{r}-\frac{r_{Q}^{2}}{r^{2}}\right)\right]
\dot{\phi}^{2}\nonumber\\
&+2ac\left(-\frac{r_{g}}{r}+\frac{r_{Q}^{2}}{r^{2}}\right)\dot{t}\dot{\phi}
=-c^2,
\end{align}
we can work out $\dot{r}$, but the expression is too tedious to be 
shown here. However, the asymptotic value looks quite simple,
\begin{align}\label{key}
\lim_{r\rightarrow+\infty}\dot{r}^{2}=c^2\left(-1+\frac{\mathcal{E}_{2}^{2}}{m^{2}}\right).
\end{align}

We assume that the probe particle starts from $r(\tau_i)=+\infty$, 
$\theta(\tau_i)=\pi/2$ 
with initial 4-velocity $u^\mu(\tau_i)=(c,0,0,0)$. These initial 
conditions imply that the integrals of motion are
\begin{align}
\mathcal{J}_{2}=0,\qquad
\mathcal{E}_{2}=mc^2. \label{E2=m}
\end{align}
Even under such initial conditions we find that the explicit evaluation 
for the works done by electromagnetic force and by gravity on a charged particle 
is extremely hard if we take a generic observer field. 
However, since $k=\partial_{t}$ and $\chi=\partial_{\phi}$ are both
Killing vector fields, it is always possible 
to choose the observer field $\mathcal{Q}$ to be static
in the region beyond the ergosphere. Thus, we shall  
restrict ourselves solely to the static observer field $\mathcal{Q}$
with $Z^\mu = \left[c\left(1-\frac{r_g r-r_Q^2}{\rho^2}\right)^{-1/2},0,0,0\right]$.

Using the above data, we are now in a position to calculate the works
$W_{\mathrm{em}}(\mathcal{Q})$ and $W_{\mathrm{grav}}(\mathcal{Q})$ 
in Kerr-Newman spacetime. In integral form, 
these works can be expressed as
\begin{align}
&W_{\mathrm{em}}(\mathcal{Q})
=\int_{+\infty}^{r_f }\frac{eQ}{4\pi\varepsilon_0 r^2}
\left(1-\frac{r_g}{r}+\frac{r_Q^2}{r^2}\right)^{-1/2}{\rm d}r,
\label{wexKN}\\
&W_{\mathrm{grav}}(\mathcal{Q})
=\int_{+\infty}^{r_f }\frac{\left(eQ-4\pi\varepsilon_0 mc^2r\right)
\left(r_gr-2r_Q^2\right)}{8\pi\varepsilon_0 r^4}\left(1-\frac{r_g}{r}
+\frac{r_Q^2}{r^2}\right)^{-3/2}{\rm d}r,
\label{wgKN}
\end{align}
provided 
\[
r_f>\frac{1}{2}\left[r_g+\sqrt{r_g^2-4r_Q^2}\,\right],
\]
i.e. the final position of the particle is located outside the
ergosphere. Clearly, the results \eqref{wexKN} and \eqref{wgKN}
coincide with eqs.\eqref{Wex for RN C=0} and \eqref{Wg for RN C=0} if 
$r_i$ were taken to be equal to $+\infty$. 
So, we conclude that, under the above choice of initial condition and observer field, 
the results for Kerr-Newman spacetime are the same as those for RN spacetime. 
The rotation of the source affects neither the work done by gravity, 
nor the work done by electromagnetic field.

\section{Concluding remarks}

To summarize, we have proposed a coordinate independent, however 
observer dependent definition for gravitational work and work done by external 
forces on a massive particle in curved spacetime. 
The definitions are then checked in Minkowski, Schwarzschild, RN and 
Kerr-Newman spacetimes with appropriate choice of observer fields, 
and their validities are justified in the far field limit. 

The result of the present paper constitutes the first step 
towards a macroscopic definition of work and heat in curved 
spacetime backgrounds. In order to achieve the macroscopic definitions
in mind, the next step may include considerations involving 
relativistic kinetic theory \cite{doi:10.1063/1.4817035,0264-9381-31-8-085013,
Cercignani2002,Hakim2011}. In this regard, the J\"uttner distribution 
\cite{doi:10.1002/andp.19113390503,Juttner1928}
may serve as a starting point. 

Alternatively, the gravitational work defined in this paper may serve 
as a tool for determining the gravitational binding energy for a 
self-gravitating system in the context of general relativity. 
The gravitational binding energy has been considered earlier in 
\cite{ADM,Israel,Binding energy for spherical stars} using different methods.
We shall come back later on this subject elsewhere.

Moreover, the relativistic rocket has long been an interesting subject 
where the gravitational work done on the rocket should be taken into 
consideration. However, the previous discussions are always restricted 
to special relativity in Minkowski spacetime. Recently, Henriques and his 
coworkers discussed about the rocket problem in curved spacetime, but they 
did not discuss about the concept of work \cite{Henriques}. Our results can 
be applied to the rocket problem in curved spacetime.

Another possible field of application of our results is in the study of 
Brownian motion and its connection to fluctuation theorems \cite{Seifert}. 
In this field, the work and free energy, as well as the non-equilibrium work relations 
are central concepts. There are already some discussions about relativistic 
Brownian motion \cite{Dunkel} and relativistic fluctuation theorems \cite{Fingerle}. 
Therefore, our results may also find applications in the study of these topics.

As a final remark, let us mention that,
in an interview with {\tt www.guokr.com}, A. Zee, a renowned 
theoretical physicist, stated \cite{AZee}, 
``Relativity is perhaps the worst
name ever in the history of physics.''  What he meant is that the 
central subject of concern in the theory of relativity is what remains 
invariant irrespective of the choice of observers, i.e. what we call 
the natural laws. This is certainly correct, but 
reflects only one side of relativistic physics.
There is another -- often neglected -- side 
of relativistic physics, i.e. the observed quantities of physical 
observables do depend on the choice of observers, albeit independent 
of the coordinate choices\footnote{An equivalent statement for this observation 
that the values of physical observables are foliation dependent \cite{Kosyakov}. 
We thank B.P. Kosyakov for letting us know about ref.\cite{Kosyakov}.}. Without this latter side, our understanding 
about relativistic physics may be incomplete. This two-sidedness of relativistic 
physics is best exemplified by the work-energy theorem \eqref{w-e int},
in which all quantities $\Delta E(\mathcal{Q}), 
W_{\mathrm{ex}}(\mathcal{Q})$ and $W_{\mathrm{grav}}(\mathcal{Q})$ are 
observer dependent, whereas the identity \eqref{w-e int} holds
for every choice of observer field $\mathcal{Q}$.

\section*{Acknowledgement}
This work is supported by the National Natural Science Foundation of 
China under the grant No. 11575088. Shaofan Liu would like to thank Xin Hao and Tao Wang for many useful discussions and suggestions.

\end{document}